# Disability in Space
## – What ESA should have aimed for –

Christiane Heinicke[1], Marcin Kaczmarzyk[2], Benjamin Tannert[3], Aleksander Wasniowski[4],
Malgorzata Perycz[5], Johannes Schöning[6]

[1]University of Bremen, Center of Applied Space Technology and Microgravity (ZARM),
Am Fallturm 2, 28 359 Bremen, Germany

[2]Rzeszow University of Technology, The Faculty of Civil and Environmental Engineering and Architecture, Department of Building Engineering, Poznańska Street 2, 35-084 Rzeszow, Poland

[3]Hochschule Bremen – City University of Applied Science, Applied Media Informatics, Flughafenallee 10, 28 199 Bremen, Germany

[4]Space is More, ul. Mikulskiego 17, 52-420 Wrocław, Poland

[5]Institute of Computer Science Polish Academy of Sciences, Department of Artificial Intelligence, Lab of Computational Biology, Jana Kazimierza 5 st., 01-248 Warsaw, Poland

[6]University of Bremen, Human-Computer Interaction, Bibliothekstraße 5, 28 359 Bremen, Germany

*ESA has announced their new "Parastronaut Feasibility Project", promising to make every reasonable effort to send astronauts with disability to space. However, the fine-print of their announcement precludes the possibility of a visionary, inspirational outcome a priori.*

On February 16[th] this year, the European Space Agency (ESA) announced for their first time in over a decade that it would be seeking new astronaut candidates. With this call, ESA will do something that is unprecedented in the entire history of human spaceflight: accept applications from individuals who have a physical disability[1]. ESA does not promise that these individuals will indeed fly to space, but assures that the whole point of the so-called Parastronaut Feasibility Project is to "attempt to clear the path" to achieve exactly that.

We very much welcome ESA's idea as a step in the right direction. An estimated 15% of the world population lives with "some form of disability" [1]. Even though only 75 million people of these live in one of ESA's 22 member states, we are thankful that ESA now provides the opportunity to these 75 million to decide for themselves if they want to apply or not. We also acknowledge that in an industry, where early astronaut candidates had to keep their feet in ice-cold water for 7 minutes [2], spend 2 hours in a room at 55℃, and only those who performed best on a two-axes rotating chair[2] would be selected – even considering candidates with physical imperfections is a giant leap forward.

Nevertheless, we argue that the current plans for the parastronaut selection are (1) far from revolutionary and hardly capable of contributing significantly to ESA's long-term visions, and (2) too focused on inclusion for the sake of inclusion, as ESA is concerned with how to include people with disabilities into the existing group of able-bodied people, instead of focusing on where people with disabilities have advantages and trying to learn from them.

### Why the current definition of the ESA Parastronaut is problematic

ESA's primary reason to include people with disabilities into their program is to comply with modern societal standards and expectations. "ESA needs and wants to embrace change in order to remain relevant and accessible [...]"[3]. In other words, ESA thinks that society wants it to include people with disabilities, and not that including people with disabilities is actually useful.

---

1 https://www.esa.int/About_Us/Careers_at_ESA/ESA_Astronaut_Selection/Parastronaut_feasibility_project
2 https://history.nasa.gov/SP-4003/ch5-5.htm

ESA selected disabilities which are expected to have the least impact in space: The only three disabilities ESA plans to accept are (1) lower limb deficiencies below the knee, (2) differences in leg length, and (3) short stature. This stands against the broader definition of "disabilities" employed by the WHO: "Disability is the umbrella term for impairments, activity limitations and participation restrictions, referring to the negative aspects of the interaction between an individual (with a health condition) and that individual's contextual factors (environmental and personal factors)" [1]. This definition clarifies that disabilities are not mere physical limitations, but are in great part barriers imposed by society.

The narrow band of "allowable" disabilities matches perfectly ESA's announcement to persuade their international partners and launch vehicle providers to make "adaptations" to existing hardware. As mentioned above, we acknowledge that starting this paradigm shift in the space industry is no easy task. Yet, sending a person with disability to space just to be able to say that one sent a person with a disability to space is as problematic as the many other attempts here on Earth where design changes often have more symbolic than practical value. Making some small adjustments after the fact hardly ever leads to useful design [3].

Let us assume that ESA succeeds in overcoming the existing reservations in the spaceflight community and sending an astronaut with a leg deficiency to orbit. The hardware adaptations are specific to this one person, this one specific type of disability and the next time a person with a different type of disability is selected, the whole process of adaptation has to be repeated. The whole approach of trying to integrate disabled people into systems that have been designed for perfectly-able-bodied people is backwards. Instead, ESA should try to learn from the experience and problem-solving genius that many people with disability have fostered over all of their lives (or since they have acquired their disability). They know their needs best and many have already found work-arounds for their specific disability.

In the longer-term perspective, making minor adjustments to existing hardware is a safe route to repeating the same inaccessible design that we have on the International Space Station and launch vehicles today in spacecraft of the future. In fact, had people with disabilities been included earlier, the design of components for Gateway could have occurred under aspects of accessibility – and made the design safer for everyone. Now that these components are already being assembled, anything to improve accessibility will again be an add-on. And the risk is that we end up with habitat on the Moon that is accessible to only a small fraction of physically perfectly-able-bodied people if ESA (and all other space agencies!) continue to make such small steps forward.

ESA is member of the International Space Exploration Coordination Group ISECG whose long-term vision is the exploration of the Moon and Mars with humans. With mission durations of 2.5 years for the latter, it is not a question if, but when, accidents will occur. The list of medical issues requiring medical attention on the ISS is long, as is the list of injuries suffered during landing on Earth after re-entry. Predictions on how much time passes between mission-impacting medical events range from 3 to 6 years for a 6-person crew [4]. On ISS, severe medical issues can be evacuated immediately. On Mars, evacuation would take several months and is therefore unlikely to be an option. Any injuries suffered on Mars will have to be coped with on site. However, Mars does not have the health care and rehabilitation facilities that we have on Earth, and thus an injured crew member must cope in the same habitat that their non-injured colleagues are using. Integrating astronauts with (permanent and well-adapted to) disabilities into today's spaceflight programs will help save a non-disabled astronaut with (temporary) injuries tomorrow.

---

3   https://www.esa.int/About_Us/Careers_at_ESA/ESA_Astronaut_Selection/Parastronaut_feasibility_project

Finally, having certain physical disabilities may actually be an advantage over starting off without any disability. What use are legs in space? Being smaller or leg-less will probably mean requiring less life support consumables, and fewer problems with adaptation to microgravity. On Mars, astronauts could move around in wheelchair-like rovers, and the space normally occupied by the legs could be used by extra life-support supplies. The blind are said to have better 3D orientation than the sighted, which would be an advantage in microgravity, and given the high rate of visual problems in space, bringing a visual impairment into space would mean no adaptation required for orientation. There are many more examples; the point is: Many physical characteristics that we perceive as "disability" on Earth may turn out to be a "hyperability" in space, and this is not limited to lower limb deficiencies.

**Dare to be more visionary**

We argue in a first step to extend the list of physical disabilities that can be accepted into the Parastronaut Feasibility Program as proposed in table 1. Moreover, we point out that certain mental disabilities are more challenging, but may help prepare better for the reduced cognitive capacity of astronauts under high stress and after long duration in space.

The next step in human exploration is the Moon and concrete preparations for it are in progress. We therefore stress that if parastronauts are not well-included now and in the program of multiple space agencies, the next spacecraft will also be built in a way so that only fit, healthy, uninjured astronauts can safely live there. Instead, it would make more sense to integrate a broader set of disabilities, bringing about completely new needs and a fresh way of thinking. **Development in space should be *with*, not *for* people with disabilities.** The benefit will be huge: Many developments with or by people with disabilities in the past have proven beneficial for people without disabilities. One of the most prominent examples is the telephone, which was a "side-product" of Graham Bell's work with the deaf[4]. More recently, Vinton Cerf might not have integrated e-mail as part of the functionality of ARPANET if he had not been hard of hearing himself.

Unlike ESA seems to believe, spaceflight is not at "step zero" in terms of inclusion. Already back in 2017 the first analog mission was conducted with a crew member with disability (ICAres-1 [5, 6]). The mission was a success; despite a combination of deformities in both hands and complete blindness, the person was a well-integrated and valued member of the crew who helped crucially in improving the safety procedures.

Lastly, change will not come only from organizations who primarily exist for the disabled. Change can be set in motion by setting an example. Spaceflight claims for itself to provide inspiration to the next generation. It is time for ESA to step up to its self-proclaimed ambition and set an example of true inspiration for the millions of people with disabilities.

*Table 1: Disabilities we propose for consideration in the Parastronaut Feasibility Project and inclusion in other spaceflight programs.*

| Type of disability | Reasoning |
| --- | --- |
| Blindness | Visual impairment is common in space (e.g., SANS, VIIP [7, 8]) |
| Deafness | Decompression sickness and barotraumata may occur after rapid decompression |

---

4   http://www.infinitec.org/history-of-tech-advances

| Type of disability | Reasoning |
| --- | --- |
| Upper leg deficiencies | Legs are not inherently needed in space |
| Hand/arm deficiencies | An astronaut with a mechanical, perhaps brainwave-driven manipulator may be superior to an able-bodied astronaut wearing EVA gloves |
| Paraplegia | Spine injury has occurred during Apollo [4] |
| Multi-morbidity | The one-disability-per-person approach would be unnecessarily time-consuming. |
| Cognitive deficiencies | Cognitive capabilities are expected to decline during long-duration spaceflight [9-11]. An astronaut with such a disability may be instrumental in devising adequate support. |
| Neurological (sensory and motor) | The central nervous system is affected by space radiation [12]. An astronaut with such a disability may be instrumental in devising adequate support. |